\documentclass{webofc}

\usepackage[varg]{txfonts}   
\usepackage{hyperref}
\usepackage{url}
\usepackage{lineno}
\hypersetup{colorlinks=true,citecolor=blue,urlcolor=blue,linkcolor=blue}
\begin{document}
%
\title{Integrating IPbus ALFRED into the ALICE-FIT setup}
%
%

\author{\firstname{Krystian} \lastname{Roslon}\inst{1}\fnsep\thanks{\email{krystian.roslon@pw.edu.pl}}
        \firstname{for the ALICE collaboration}
}

\institute{Faculty of Physics, Warsaw University of Technology, \\ Koszykowa 75, 00-662 Warsaw, Poland
          }

\abstract{The official data collection for the Run 3 of the Large Hadron Collider (LHC) at CERN in Geneva commenced on July 5, 2022, following approximately three and a half years of maintenance, upgrades, and commissioning. Among the many upgrades to ALICE (A Large Ion Collider Experiment) is the new Fast Interaction Trigger (FIT) detector. Constant improvements to FIT's hardware, firmware, and software will enable progressively better performance. Between November 2024 and March 2025, during the year-end technical stop, an update to the communication path between the Front-End Electronics (FEE) and the Detector Control System (DCS) is planned. This update will introduce a new approach based on the ALFRED (ALICE Low-Level Front-End Device) software, supported by the central DCS ALICE system. To address the challenge of integrating custom electronics with distributed control systems, this paper describes a novel extension of the Front-End Device (FRED) framework, which can interface bespoke electronics with standard SCADA (Supervisory Control and Data Acquisition) systems using IPbus. This framework can be applied to all detectors utilizing IPbus communication. }
\maketitle
\section{Motivation and concept}

The ALICE \cite{Aamodt:2008zz} experiment, conducted at CERN \cite{cernpage}, is a prime example of a large-scale research facility that is heavily influenced by the diverse range of technologies utilized by participating institutes across different countries. For smaller installations, industrial standards such as OPC (Open Platform Communications) \cite{OPC} are employed to ensure a consistent interface between high-level SCADA systems and underlying electronics. However, in the case of ALICE, the integration of hundreds of modules with unique functionalities into a single control system poses a significant challenge. Despite this, the unified Detector Control System, utilizing the WinCC OA \cite{WinCCOA} standard SCADA system, is responsible for managing all detectors of the ALICE experiment. This system must be user-friendly and capable of being operated by a single individual. The WinCC OA system, developed by Siemens, is the standard SCADA/HMI system used for all detectors in the ALICE experiment. Additionally, the FRED \cite{Jadlovsky:2018tux} framework serves as a new software layer within the control architecture, facilitating the connection between custom detector electronics and the standard SCADA system, while also providing a high level of abstraction to eliminate the need for the SCADA system to have knowledge about the specific type of electronics being used.

\begin{figure}[!ht]
    \begin{center}
        \includegraphics[scale=0.4]{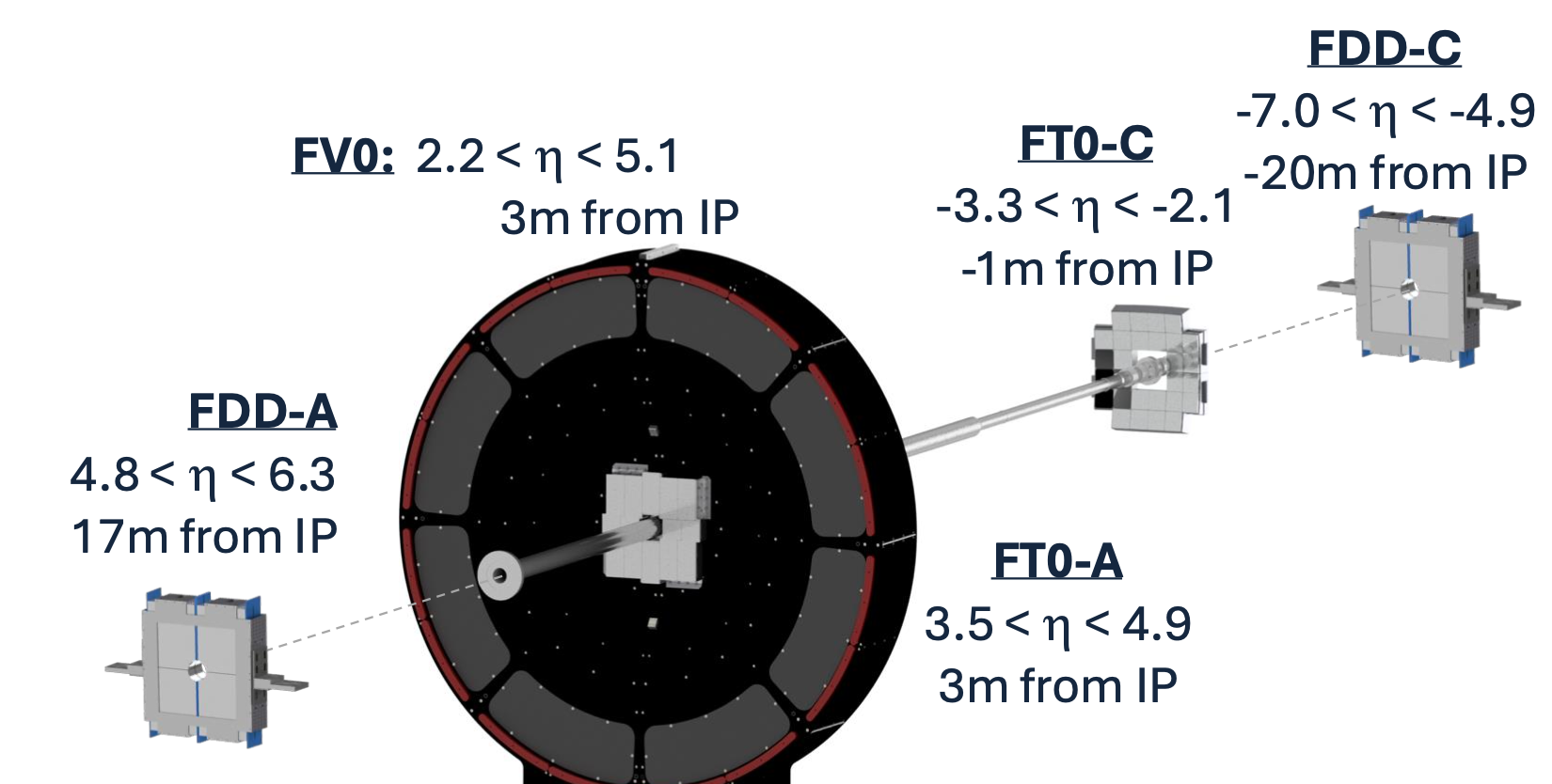}
    \end{center}
    \caption{Layout of FIT detector with the location along the beamline and the pseudorapidity coverage of each subdetector array.}
    \label{fig:FIT}
\end{figure}

Among the many upgrades to ALICE is the new Fast Interaction Trigger detector. FIT consists of three distinct detectors: FT0, FV0, and FDD. These detectors utilize both Cherenkov radiation (FT0) and scintillation (FV0 and FDD) phenomena to register charged particles produced in proton-proton (pp) and heavy-ion collision events \cite{Slupecki:2022fch}. The schema of the detector illustrated in Fig. \ref{fig:FIT}.

\begin{figure}[!ht]
    \begin{center}
        \includegraphics[scale=0.45]{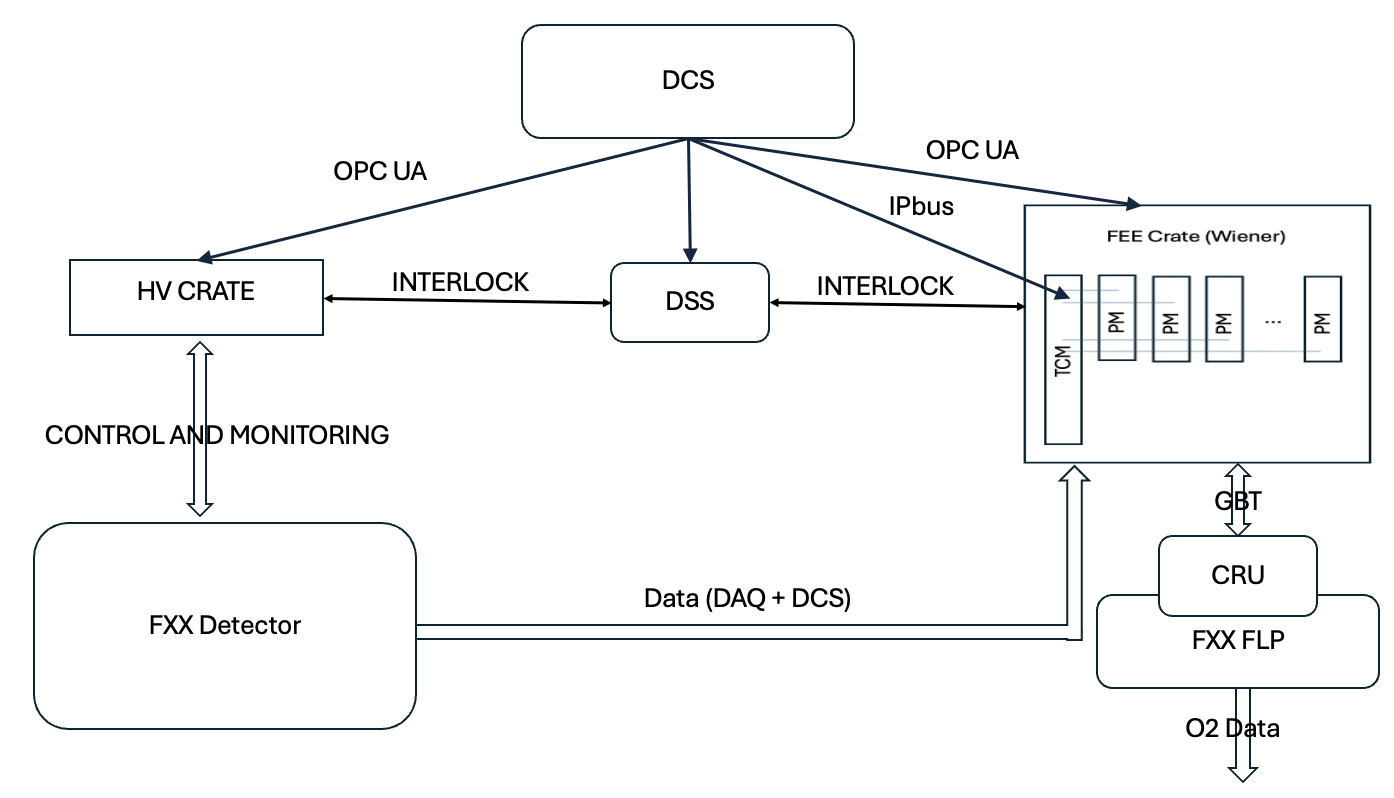}
    \end{center}
    \caption{Schematic representation of the Detector Control System for the FIT detectors, where FXX: defines the detector -  FT0/FV0/FDD.}
    \label{fig:DCS_SCHEMATIC}
\end{figure}

The architecture of the recently developed FIT control system replicates the structure of the detector illustrated in Fig. \ref{fig:DCS_SCHEMATIC}. Access to the FEE is facilitated through the GBT link \cite{Antonioli:2013ppp}, which is shared with the data acquisition system. The physics data, encompassing information regarding the interaction trigger, online luminometer, initial indicator of the vertex position, and the forward multiplicity counter, is transmitted to the subsequent processing stages executed at the O2 facility \cite{Konopka:2020ten}.

In addition to the front-end control, the primary DCS components encompass the control system for HV (CAEN) and FEE (Wiener) crates. 

\begin{figure}[!ht]
    \begin{center}
        \includegraphics[scale=0.35]{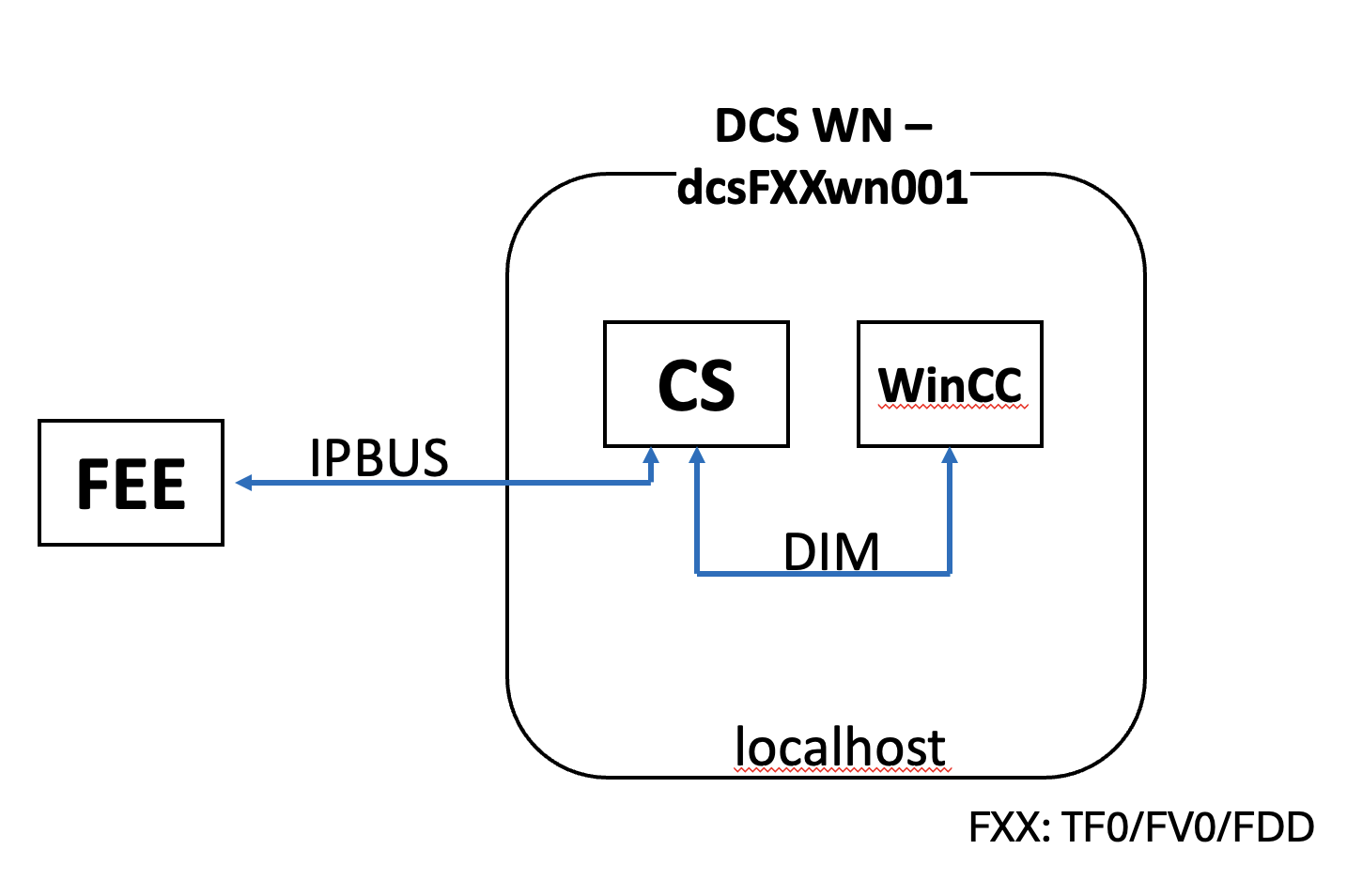}
    \end{center}
    \caption{Actual status of Detector Control System for the FIT setup, CS (ControlServer) and WinCC are located on the same machine called ``WorkerNode'' (WN) where FXX: defines the detector - FT0, FV0, FDD. The comunication with the FEE bases on IPbus protocol.}
    \label{fig:CSFLOW}
\end{figure}

At present, the interaction between WinCC OA and FEE involves the utilization of a custom software called ``ControlServer''  designed espacially for the FIT detector needs. This software employs IPbus protocol \cite{Larrea_2015} to establish communication with the electronics. Both the WinCC OA software and the ``ControlServer'' are deployed on a single machine, facilitating the exchange of DIM \cite{Gaspar:2001fbw} (Distributed Information Management System) services and commands locally (on the same machine) to support the communication process (see Fig. \ref{fig:CSFLOW}).

This solution is distinctive as it diverges from the conventional approach employed by other ALICE subdetectors, which utilize communication through GBT links in conjunction with Front-End Electronics. In contrast, the FIT system communicates via IPbus. Furthermore, the aforementioned solution lacks support from the central ALICE Detector Control System team. Consequently, it is imperative to integrate the DCS for FIT with the ALICE central DCS and to implement the ALFRED framework \cite{Chochula:2018vfx}.

\section{ALFRED implementation in ALICE}
\begin{figure}[!ht]
    \begin{center}
        \includegraphics[scale=0.28]{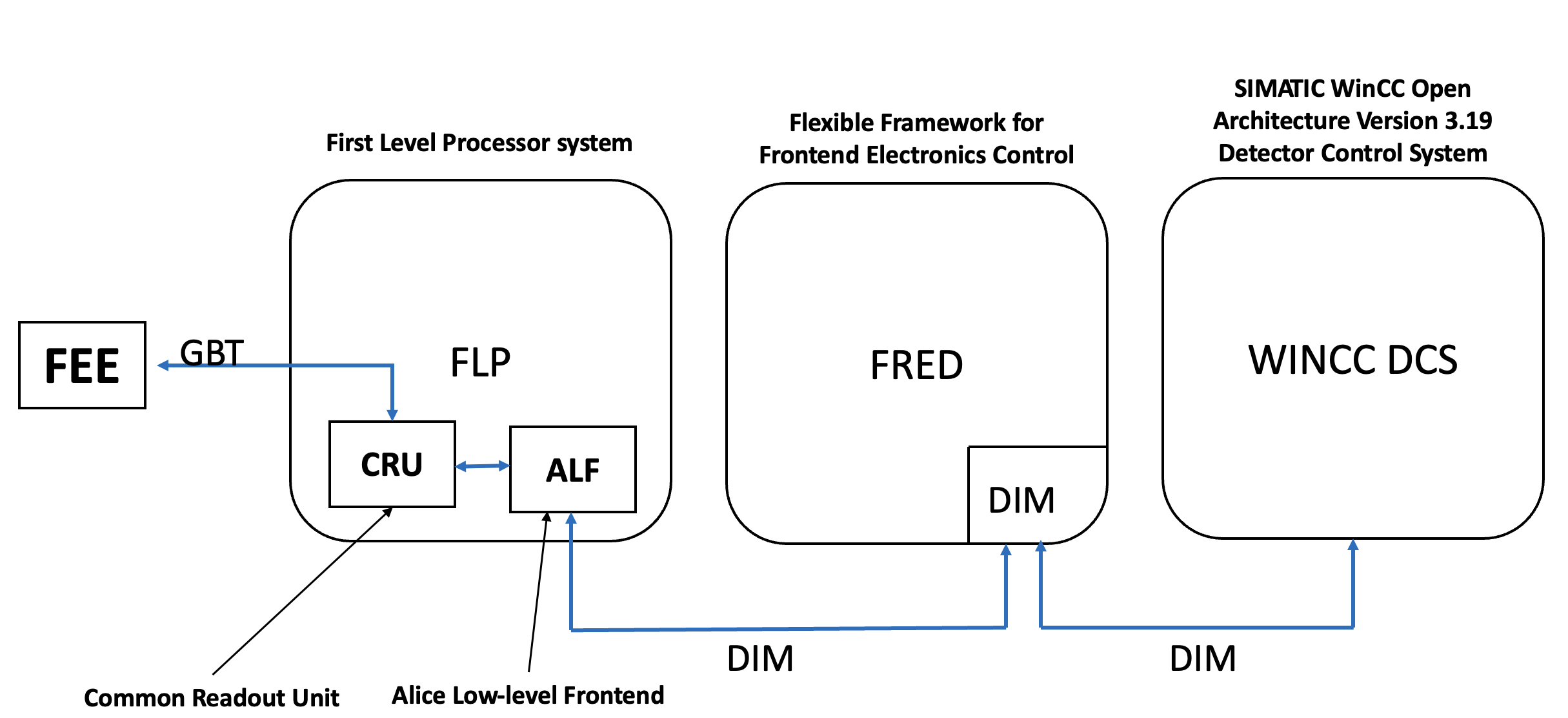}

    \end{center}
    \caption{Final solution for the Detector Control System based on the GBT communication with the FEE for the FIT setup which is supported by the central ALICE-DCS team.}
    \label{fig:GBT_FLOW}
\end{figure}

ALFRED is a framework that consists of two modules ALF (ALICE low-level front-end) and FRED. It establishes communication with detector electronics through ALF server applications that are executed on FLP computers. ALF serves as a straightforward intermediary that enables remote processes to access the CRU (Common Readout Unit) driver which reaches the FEE via GBT links. ALF itself does not execute any data operations; rather, it acts as a bridge between the DIM network protocol and the detector electronics. Additionally, it guarantees the atomicity of driver access, thereby ensuring the transmission of command sequences as a unified entity and averting conflicts in scenarios where multiple remote processes make simultaneous requests (see. Fig. \ref{fig:GBT_FLOW}). It is compatible with the FRED program module, which is responsible for gathering data from ALF modules and transmitting it to the WinCC OA visualization tool. Additionally, it retrieves configuration data from the database and transmits it to ALF modules through the TCP protocol. Moreover, the FRED module incorporates the DIM command which transmits the parameters to the front-end electronics via ALF modules. \cite{Jadlovsky:2018tux}

The approach mentioned above is implemented in various sub-detectors of the ALICE experiment and is also supported by the central ALICE-DCS system. It represents the standard solution adopted by the ALICE-FIT collaboration. Nevertheless, the existing firmware programmed in the TCM (Trigger and Clock Module) and the PMs (Processing Module) FPGA of the FIT detector electronics, which operate through IPbus-based communication, necessitates modifications to the ALFRED framework to accommodate this form of communication.

\section{IPbus integration in ALFRED}

In the field of high-energy physics experiments, the establishment of accurate and dependable communication between control systems and hardware is of paramount importance. The IPbus protocol, specifically designed to address these rigorous requirements, emerges as a critical element in the configuration and oversight of hardware utilized in particle physics investigations.

IPbus acts as a communication protocol that connects host computers with various electronic modules, including FPGA-based devices, within the data acquisition systems of particle physics experiments. Its core responsibilities encompass configuring hardware registers, monitoring the operational status of hardware components, and relaying vital control commands. A significant advantage of IPbus is its functionality over standard TCP/IP networks. This compatibility with widely used network infrastructures facilitates flexible and scalable implementations across diverse experimental configurations. By utilizing established network technologies, IPbus guarantees smooth integration into a variety of settings.

Communications within IPbus are transaction-oriented, consisting of a client request followed by a corresponding server response. This approach ensures organized and dependable exchanges. The protocol utilizes a defined packet format that includes information such as the transaction type (e.g., read or write), the address of the target register, and the data to be either written or read. This standardized format streamlines the development and upkeep of communication processes \cite{Larrea_2015}.

\begin{figure}[!ht]
    \begin{center}
        \includegraphics[scale=0.28]{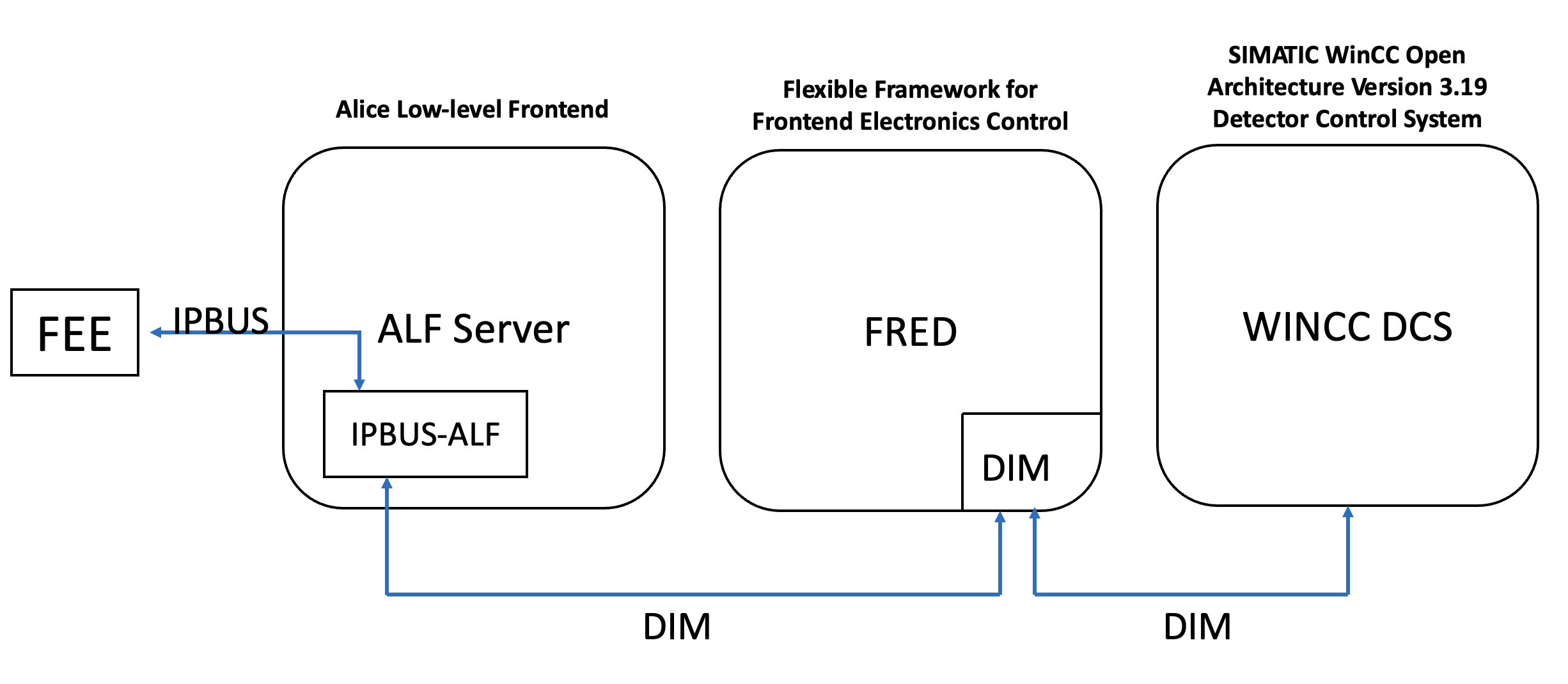}
    \end{center}
    \caption{Detector Control System based on the IPbus communication with the FEE for the FIT setup.}
    \label{fig:IPBUS_FLOW}
\end{figure}

At first glance, the most effective approach appears to be linking the FEE directly to the FLP and deploying an application that interfaces through IPbus on that specific machine. Nevertheless, it is important to note that the FLP operates on a distinct network separate from the DCS servers, which encompass the ``WorkerNode'' and the FRED server. Furthermore, ALF is integrated within O2 and interacts with the FEE through a GBT connection. This scenario necessitated the development of a solution that could be executed without altering the FPGA firmware (see Fig. \ref{fig:IPBUS_FLOW}).

The proposed solution delineates the division of the DCS across a 3 machines. In accordance with the standard configuration, the SCADA system, developed using WinCC OA, runs on a designated machine referred to as ``WorkerNode''. Communication between the ``WorkerNode'' and the FRED server utilizes DIM services and DIM commands. Following the completion of calculations, the resultant data is transmitted to the ALF server, which executes the ALF-like (IPbus-ALF) program. The interaction between FRED and IPbus-ALF is also facilitated through DIM. The primary function of the IPbus-ALF program is to relay data to the FEE via the IPbus protocol, circumventing the CRU. Consequently, this program is installed on a dedicated machine rather than the FLP. All systems are interconnected within a local, secure network that includes the FEE.

\section{Conclusion and Further Works}

The objective of this article is to present a proposed communication architecture for the Detector Control System tailored for the Fast Interaction Trigger and to describe the potential application of the IPbus protocol for communication with the Front-End Electronics. This solution is also applicable to other detectors utilizing the IPbus protocol. Consequently, it is essential to meet and consider complex conditions. The architecture is based on several distinct implementations, including WinCC OA, FRED, ALF, DIM, and IPbus. Future research will focus on replacing the IPBus-ALF structure with a system that translates the SWT \cite{Bourrion:2019lev} protocol into IPbus. As a result of this work, a plug-and-play system will be developed, capable of communicating with the Control Readout Unit and testing the bandwidth of the modified architecture.

\section{Acknowledgments}
This work was supported by the Polish Ministry for Education and Science under agreements no. 5452/CERN/2023/0.
\bibliography{bibliography}

\end{document}